# Characterization of the recovery of mechanical properties of ion-implanted diamond after thermal annealing

M. Mohr[1], F. Picollo[2,3,4], A. Battiato[3,2,4], E. Bernardi[3,2,4], J. Forneris[3,2,4],

A. Tengattini[3,2,4], E. Enrico[5], L. Boarino[5], F. Bosia[3,2,4], H.-J. Fecht[1], P. Olivero[3,2,4,5]*

[1] *Institute of Micro and Nanomaterials, Ulm University, Ulm, Germany.*

[2] *Istituto Nazionale di Fisica Nucleare (INFN), Section of Torino, Italy.*

[3] *Physics Department and "Nanostructured Interfaces and Surfaces" Inter-departmental Centre - University of Torino, Torino, Italy.*

[4] *Consorzio Nazionale Inter-universitario per le Scienze Fisiche della Materia (CNISM), Section of Torino, Italy.*

[5] *Istituto Nazionale di Ricerca Metrologica (INRiM), Torino, Italy.*

\*　　　　corresponding author: paolo.olivero@unito.it

**Abstract**

Due to their outstanding mechanical properties, diamond and diamond-like materials find significant technological applications ranging from well-established industrial fields (cutting tools, coatings, etc.) to more advanced mechanical devices as micro- and nano-electromechanical systems. The use of energetic ions is a powerful and versatile tool to fabricate three-dimensional micro-mechanical structures. In this context, it is of paramount importance to have an accurate knowledge of the effects of ion-induced structural damage on the mechanical properties of this material, firstly to predict potential undesired side-effects of the ion implantation process, and possibly to tailor the desired mechanical properties of the fabricated devices. We present an Atomic Force Microscopy (AFM) characterization of free-standing cantilevers in single-crystal diamond obtained by a FIB-assisted lift-off technique, which allows a determination of the Young's modulus of the diamond crystal after the MeV ion irradiation process concurrent to the fabrication of the microstructures, and subsequent thermal annealing. The AFM measurements were performed with the beam-bending technique and show that the thermal annealing process allows for an effective recovery of the mechanical properties of the pristine crystal.

## 1. Introduction

MeV ion implantation has been widely exploited in recent years for the micro-fabrication of single-crystal diamond, through the implementation of the so-called "lift-off technique" [1-4]. This technique can be effectively adopted to fabricate micro-mechanical structures in single-crystal diamond, with applications ranging from high-frequency MEMS devices [5-11] to opto-mechanical resonators [12], thus taking advantage of the extreme mechanical properties of diamond [13]. Recently, the latter topic attracted significant interest due to the outstanding properties of nitrogen-vacancy centers in diamond [13], whose spin-dependent optical transition can effectively couple with local mechanical stresses [14-16]. To this end, various different techniques have been employed to fabricate opto-mechanical resonators in diamond [17-20].

In the case of the lift-off technique, the fabrication process is based on the local conversion of diamond to a sacrificial graphitic layer through MeV-ion-induced damage and subsequent thermal annealing [4]. The fabrication technique is very versatile, because the local induced damage density can be controlled by varying implantation parameters (namely, ion energy, species and fluence). Nonetheless, a residual damage density (and related mechanical stress) is induced in the non-sacrificial regions as a side-effect of the fabrication technique [21]. Similarly, with other fabrication techniques [17-20], a residual damage can be induced in the fabricated opto-mechanical microstructures, particularly if ion implantation is adopted to induce the formation of nitrogen-vacancy centers [22]

For these reasons, it is necessary to accurately estimate deformation and stress levels to reliably design and fabricate MEMS structures. Moreover, the variation of elastic properties of damaged diamond as a function of induced damage density and post-processing (annealing) parameters remains to be clarified. In particular the Young's modulus of ion-implanted diamond can potentially vary between that of pristine diamond (>1 TPa, in the presence of no damage) to that of amorphous carbon (~10 GPa, for full amorphization), i.e. over two orders of magnitude. Clearly, this large variation in elastic properties is likely to strongly affect modelling results in the fabrication of mechanical and opto-mechanical resonators. Attempts have been made to experimentally derive the variation of elastic properties of diamond as a function of induced damage, but only indirect estimations with limited accuracy have been obtained [23]. This lack of experimental data is partly due to its high Young's modulus, which makes it difficult to perform indentation experiments. Here, we perform a study of the elastic properties of ion-implanted and subsequently annealed single-crystal diamond by means of Atomic Force Microscope (AFM) measurements on free-standing cantilever structures microfabricated with a FIB-assisted lift-off technique [3, 4].

## 2. Micro-fabrication

An artificial diamond sample grown by High Pressure High Temperature (HPHT) by ElementSix (UK) was employed in this work (product code: 145-500-0040). The sample is 2.6×2.6×0.5 mm$^3$ in size and is classified as type Ib, on the basis of a nominal concentration of substitutional nitrogen <200 ppm. As indicated by the producer, the sample typically consists of ~80% single <100> sector, thus the effects of growth-sector boundaries on its mechanical properties are negligible. It is known that HPHT diamond samples contain catalyst impurities such as iron, nickel or cobalt, typically at concentration levels of the order of ~10 ppm [24-26]. At such concentrations, the effect of metallic impurities on the mechanical properties is negligible, and the Young's modulus of these "mechanical grade" crystals is comparable to that of CVD-grown samples (i.e. ~1.1 TPa [27]). The sample is cut along the [100] crystal direction and is optically polished on one of the two opposite large faces. The sample was implanted at room temperature across one of the polished surfaces with 800 keV He$^+$ ions at the AN2000 accelerator of the INFN National Laboratories of Legnaro with a focused ion beam, in order to deliver a fluence of $1\times10^{17}$ cm$^{-2}$. The microbeam spot was ~10 μm in diameter, and was raster-scanned to implant a rectangular area of ~500×200 μm$^2$. The high density of damage induced by ion implantation promotes the conversion of the diamond lattice into an amorphous phase within a layer which is located at ~1.4 μm below the sample surface, as shown in Fig. 1.

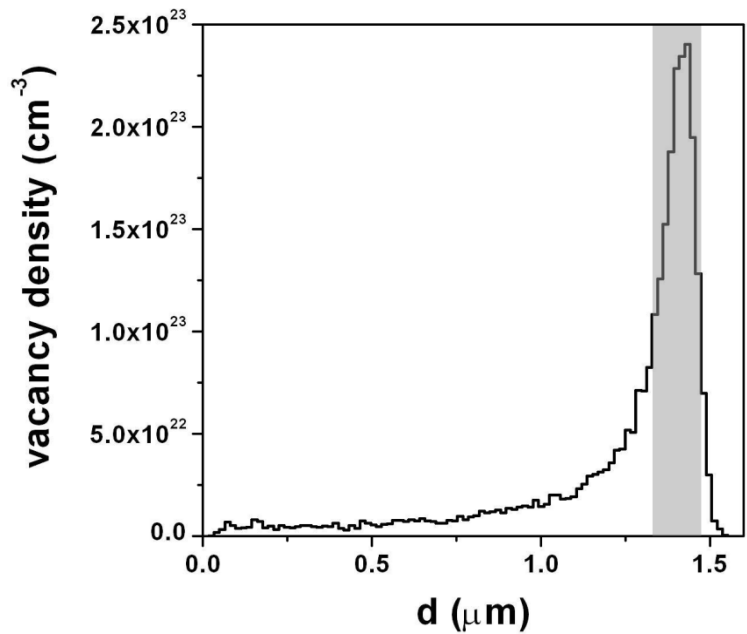

*Fig. 1: depth profile of the vacancy density induced by 800 keV He$^+$ ion implantation at fluence $1\times10^{17}$ cm$^{-2}$, as evaluated with SRIM2013.00 Monte Carlo code [28] and assuming a linear dependence of the induced vacancy density from the implantation fluence [29].*

The sample was then annealed in high vacuum (~10$^{-6}$ mbar) at 1000 °C for 1 h, to convert the highly-damaged regions located at the ion end of range to a graphitic phase while removing the structural sub-threshold damage introduced in the layer overlying the damaged region. Following the fabrication scheme described in [1], FIB milling with 30 keV $Ga^+$ ions was subsequently performed on the implanted area, to expose the sub-superficial graphitic layer to the subsequent etching step, while defining the geometries of three different cantilever structures characterized by different widths (see Fig. 2). A thin Au film was deposited on the sample surface to avoid charge effects during FIB micro-machining. The sample was then exposed to contact-less electrochemical etching [30]: the sample was placed for several hours in de-ionized water with the region of interest comprised between two close (i.e. few millimeters) Pt electrodes kept at a DC voltage difference of ~100 V. The process resulted in the selective removal of the sacrificial graphitic layer and in the creation of free-standing cantilever structures with a lateral geometry defined by the previous FIB micromachining, i.e. a length of 117 μm for cantilevers #1 and #2 and of 111 μm for cantilever #3. The widths of cantilevers #1, #2 and #3 were respectively 13 μm, 9 μm and 22 μm. The thickness of all cantilevers corresponded to the penetration depth of the employed 800 keV ions, i.e. 1.3 μm. As shown in Fig. 2, all cantilevers are slightly bent upwards due to residual stresses induced by the fabrication process within the "cap layer" comprised between the sub-superficial graphitic layer and the sample surface. In previous works this effect has been observed in structures fabricated with the lift-off method [31] and is due to the inhomogeneous swelling through the depth of the beam caused by the non-uniform ion damage profile, as shown in Fig. 1. Thermal annealing is known to reduce this effect, but it does not remove entirely this effect of residual defects. This material swelling at the underside of the beam can also cause a slight local increase in the beam cross section, however this effect is negligible with respect to the FIB-machining related effects, which are discussed below.

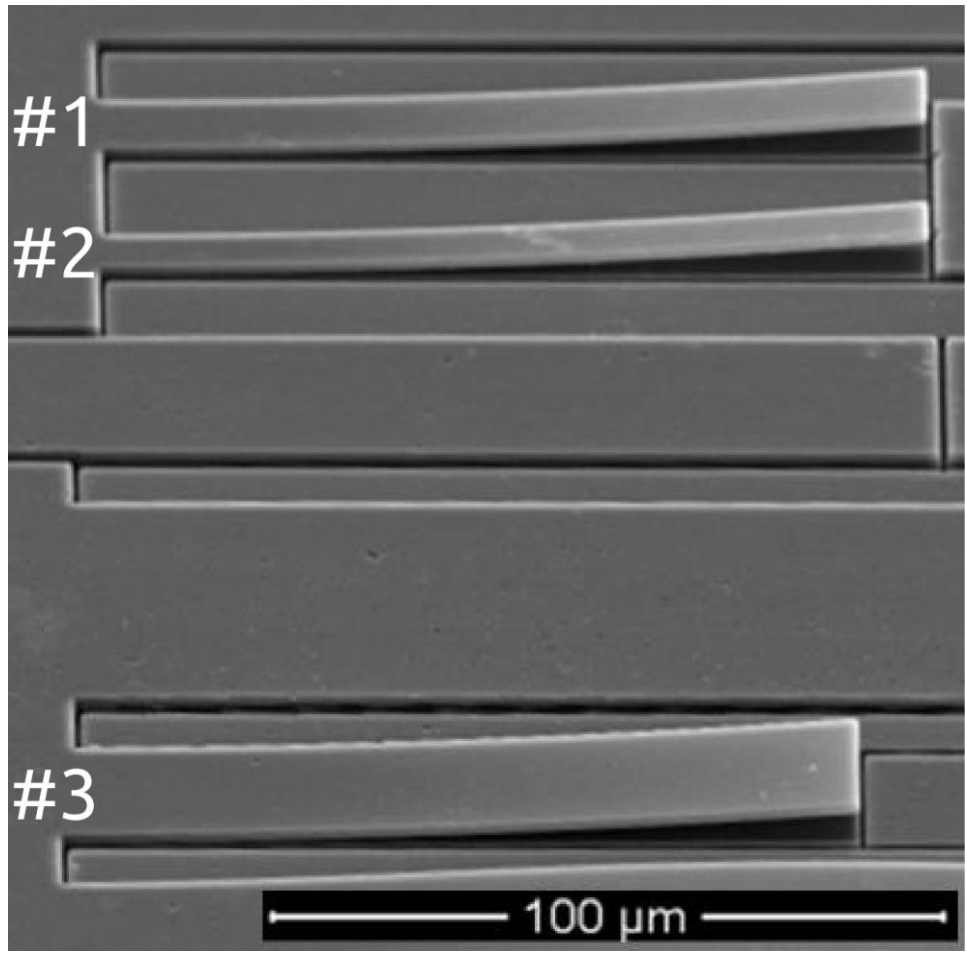

*Fig. 2: SEM micrograph of the three free-standing cantilever structures fabricated in single-crystal diamond by means of the FIB-assisted lift-off technique.*

**3. AFM Characterisation**

In order to determine the Young's modulus of the diamond, a beam-bending method is employed [32-34]. The method consists in loading the microstructures using an AFM cantilever. As shown in Fig. 3, the deflection *d* of the probing AFM cantilever for the displacement *z* of the piezomotor is measured by means of a laser diode and a position-sensitive photodiode (Veeco Dimension 3100).

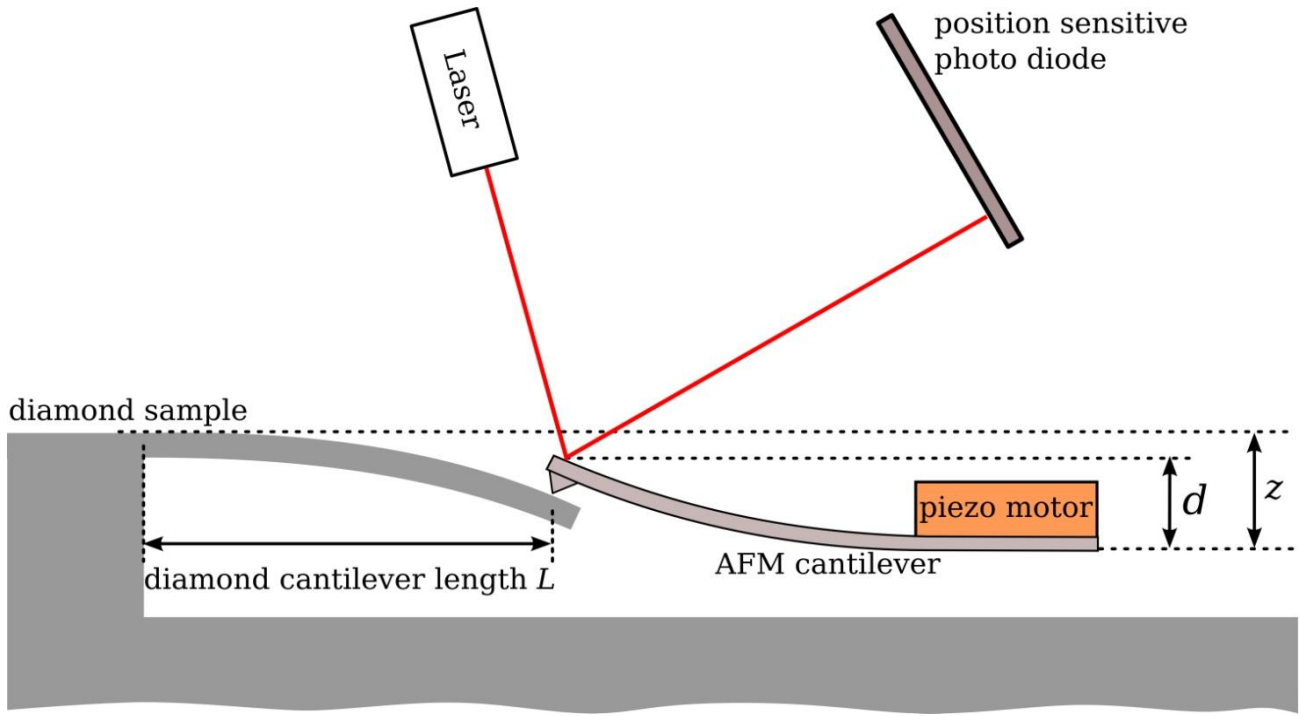

*Fig. 3: Schematic representation (not to scale) of the beam-bending technique employed to measure the Young's modulus of the microfabricated diamond cantilevers. An AFM cantilever loads the diamond microstructure at a length L, while the displacement d of the AFM cantilever is measured as a function of the piezomotor displacement z.*

The effective stiffness $k_{eff}$ of the system based on the coupling of the two cantilever structures is measured by recording approach curves. The effective stiffness $k_{eq}$ is equal to:

$$\frac{\overline{\quad\quad}}{\overline{\quad}\cdot\overline{\quad}} \tag{1}$$

where $k_{AFM}$ and $k_{diam}$ are respectively the stiffness values of the probing AFM cantilever and of the diamond cantilever under test. The $d/z$ value is equal to [32]:

$$\overline{\quad} - \overline{\quad} \tag{2}$$

Thus, the stiffness of the beam under test can be determined, if the stiffness of the AFM cantilever is known. The adopted AFM cantilever is a single-crystalline silicon cantilever (NCLR, Nanoworld). Its geometry is precisely determined by means of SEM microscopy and its stiffness was evaluated as $k_{AFM} = (57.0 \pm 1.2)$ N m$^{-1}$.

The three diamond cantilevers mentioned above are investigated, and for each of them the stiffness values are measured in correspondence of several (i.e. >5) different positions along the cantilever axis. The stiffness of the diamond cantilever is determined using the following formula:

$$\frac{(\quad)}{\quad} \qquad (3)$$

where $E$ is its Young's modulus, $I$ is its areal moment of inertia, $v$ is its Poisson's ratio, and $L$ is its length, i.e. the distance from the clamping point where the load is applied [35]. Due to a possible systematic error in the determination of the cantilever length and the finite stiffness of the cantilever fixture, we used the following correction to fit the $k_{diam}$ vs $L$ trend [36, 37]:

$$\frac{(\quad)(\quad)}{\quad} \qquad (4)$$

The Poisson's ratio is estimated using the value corresponding to pristine undamaged diamond, i.e. 0.105 [38], considering bending along the (100) direction, consistently with the well-defined orientation of the structure with respect to the crystal orientation. The geometry of the diamond beam is measured by means of SEM microscopy. As shown in Fig. 4, the cross-section appears to be slightly trapezoidal, due to a well-known effect in FIB milling related to the re-deposition of milled material that leads to a tapering of the cuts in the cross-sectional direction. Therefore, the moment of inertia I of each cantilever is calculated by taking into consideration its trapezoidal cross section (see Fig. 4), as follows [39]:

$$\frac{(\quad\quad)}{(\quad)} \tag{5}$$

where $h$ is the thickness of the cantilever and $b_1$ and $b_2$ are the two measured widths of the trapezoidal cross section (see Fig. 4). The Young's modulus $E$ of the diamond cantilever is then calculated using by fitting the $k_{diam}$ vs $L$ trend with Eq. (4), in which the moment of inertia $I$ is estimated as reported in Eq. (5).

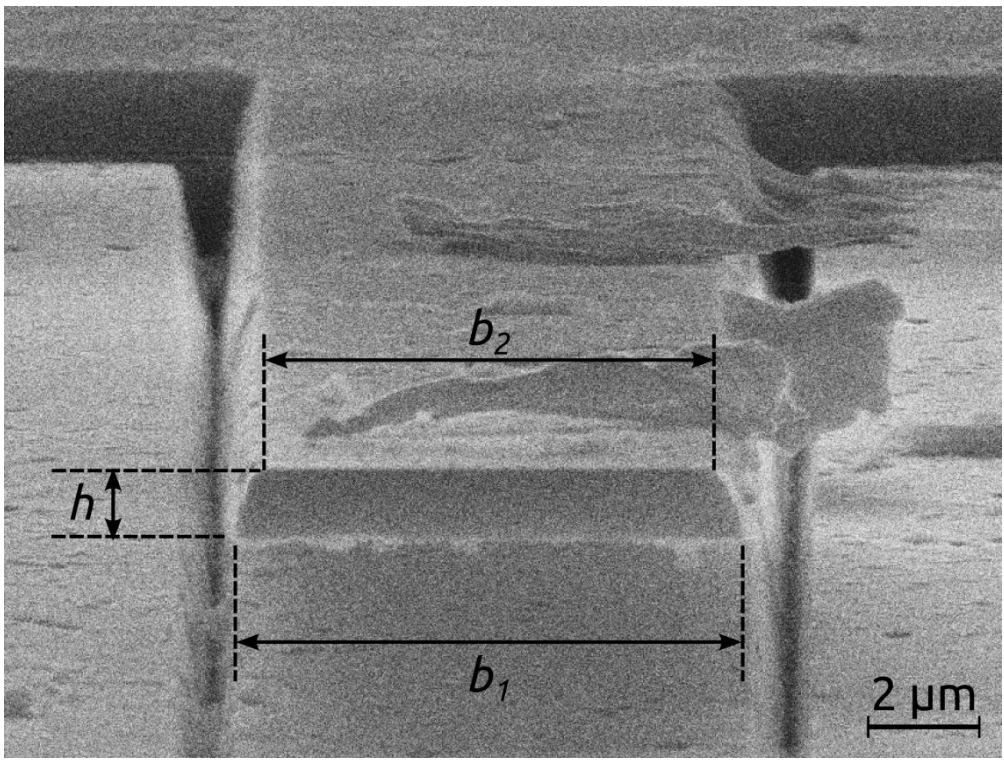

*Fig. 4: SEM micrograph of the trapezoidal cross-section of cantilever #2. h is the thickness of the cantilever, while $b_1$ and $b_2$ are respectively the lower and upper widths of the cross section.*

Representative results of the $k_{diam}$ vs $L$ measurements relative to cantilever #3 are reported in Fig. 5a, together with the fitted curve (see using Eq. (4)). The fit of the experimental data is very satisfactory, yielding a Young's modulus value of $E = (9.6 \pm 1.1) \times 10^2$ GPa. The Young's moduli for the three cantilevers are reported in Fig. 5b, together with their weighted average value and its relevant uncertainty, as well as with a comparison with the reference value of pristine single-crystal diamond. In the case of cantilever #1, the discrepancy between the measured Young's modulus and the reference value is statistically significant, and the calculated stiffness value is exceedingly high. This can be tentatively attributed to a non-ideal detachment of the beam from the substrate during the etching fabrication process, which potentially increases the stiffness of the structure and yields incorrect values when using the above formulas. Nevertheless, the weighted average of the three Young's modulus values yields an estimation of $E = (1.11 \pm 0.08)$ TPa, which is in excellent agreement with the reference value in literature for single-crystal diamond, i.e. 1.13 TPa [27]. This compatibility is remarkable, particularly if we consider that we assumed a value of the Poisson's ratio corresponding to the pristine diamond, thus introducing a potential systematic error.

Since single-crystal diamond is mechanically anisotropic [38], with, the Young's modulus in the (100) direction can be calculated as follows:

$$E_{(100)} = \frac{1}{S_{11}} \tag{6}$$

where $S_{11}$ is the 11 component of elastic coefficient tensor $S_{ij}$ according to the standard notation. Thus, the Poisson's ratio describing the contraction in the (001) direction due to tension in the (100) direction can be calculated as follows:

$$\nu_{(100)} = -\frac{S_{12}}{S_{11}} \tag{7}$$

From the values of the $S_{ij}$ elastic coefficients reported in [41], we obtain $E_{(100)} = 1.05$ TPa and $\nu_{(100)} = 0.11$. Again, these values are close to the literature values for single-crystal diamond [40].

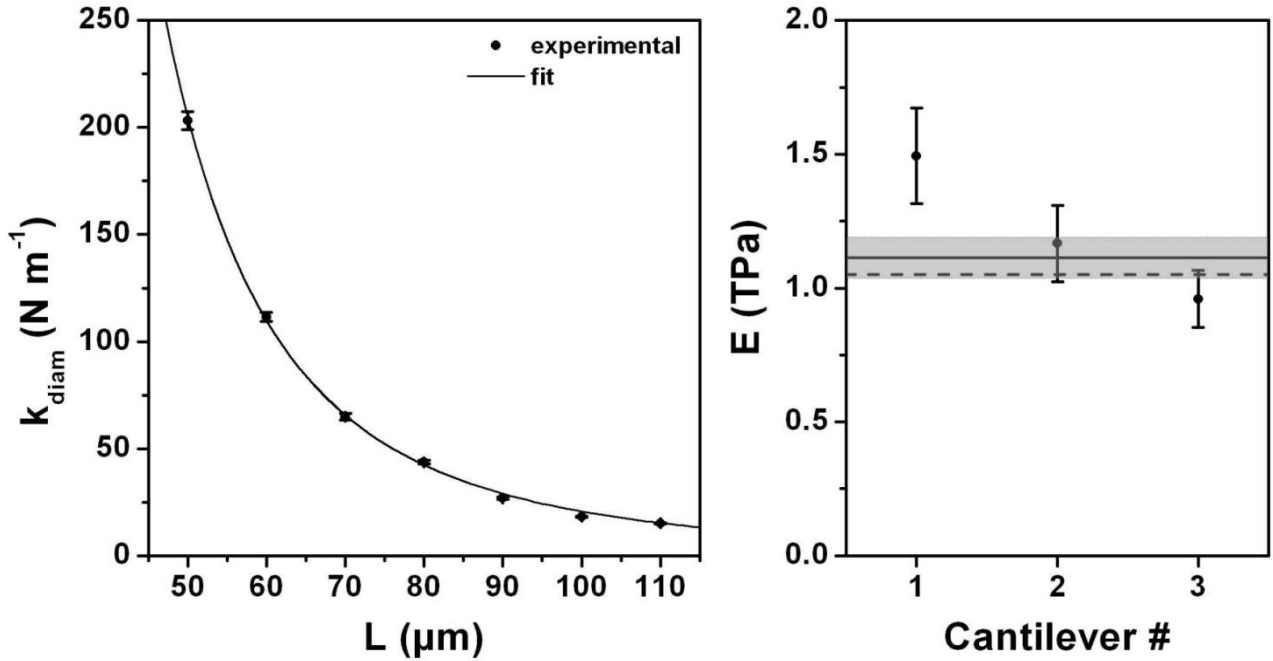

*Fig. 5: a) experimental values (dots) of the stiffness of cantilever #3 at different beam lengths together with the relevant fitting curve (line). b) Young's moduli for the three cantilevers estimated from the fit of the experimental data (dots), together with the weighted average value (solid line) and the relevant uncertainty (grey box); the Young's modulus value for pristine single-crystal diamond taken from literature [40] is reported for comparison (dashed line).*

## 4. Conclusions

We demonstrated the feasibility and reliability of an AFM-based beam-bending technique to determine the mechanical properties single-crystal diamond cantilevers, and investigated the effects of MeV ion implantation and subsequent high-temperature annealing on these mechanical properties. The obtained results provide direct evidence that both the Young's modulus and Poisson's ratio values of diamond after sub-graphitization-threshold irradiation and high-temperature annealing are fully recovered to their pristine values. These results provide useful information for the reliable design of (opto-) mechanical resonators in single-crystal diamond. We envisage to repeat these measurements on micro-structures subjected to controlled ion irradiation, with the purpose of directly investigating the effect of ion-induced damage on their mechanical properties, thus allowing the fine-tuning of their resonance frequencies.

## Acknowledgements

This work was supported by the following projects: "DiNaMo" (young researcher grant, project n° 157660) by INFN; FIRB "Futuro in Ricerca 2010" (CUP code: D11J11000450001) funded by MIUR and "A.Di.N-Tech." (CUP code: D15E13000130003) funded by the University of Torino and "Compagnia di San Paolo". The MeV ion beam lithography activity was performed within the "Dia.Fab." experiment of the INFN Legnaro National Laboratories. The FIB lithography was performed at the "NanoFacility Piemonte" laboratory, which is supported by the ''Compagnia di San Paolo'' Foundation.